\documentclass[aps, preprint, prd]{revtex4-2}

\usepackage{graphicx}
\usepackage{multirow}
\usepackage{amsmath}
\usepackage{amstext}
\usepackage{amssymb}
\usepackage{array}
\usepackage{hyperref}
\usepackage{enumerate}
\usepackage{xspace}

\usepackage{wasysym}
\usepackage[dvipsnames]{xcolor}
\usepackage{here}
\usepackage{epsfig}
\usepackage{epstopdf}
\usepackage{soul}
\usepackage{float}
\usepackage{bigints}
\usepackage{ulem}
\usepackage{xcolor}

\newcommand{\comma}{{\quad , \quad}}

\renewcommand{\d}[1]{\mathinner{d#1}}
\newcommand{\fn}[2]{\mathinner{#1\mathopen{\left(#2\right)}}}
\newcommand{\eq}[1]{Eq.~(\ref{#1})}
\newcommand{\eqs}[2]{Eqs.~(\ref{#1}) and (\ref{#2})}

\newcommand{\eqsss}[4]{Eqs.~(\ref{#1}), (\ref{#2}), (\ref{#3}) and (\ref{#4})}

\newcommand{\Rm}{\mathcal{R}_{\delta\rho_\mathrm{m}}}
\newcommand{\Rr}{\mathcal{R}_{\delta\rho_\mathrm{r}}}

\newcommand{\GeV}{\mathinner{\mathrm{GeV}}}

\newcommand{\invMpc}{\mathinner{{\rm Mpc}^{-1}}}

\newcommand{\Fref}[1]{FIG.~\ref{#1}}

\begin{document}

\title{Modeling Cosmological Perturbations of Thermal Inflation }

\author{Jeong-Myeong Bae}
\affiliation{Center for Theoretical Physics, Department of Physics and Astronomy, Seoul National University, Seoul 08826, Republic of Korea}
\affiliation{School of Undergraduate Studies,
College of Transdisciplinary Studies, DGIST,
Daegu 42988, Republic of Korea}

\author{Sungwook E. Hong}
\affiliation{Korea Astronomy and Space Science Institute, Daedeok-daero 776, Yuseong-gu, Daejeon 34055, Republic of Korea}
\affiliation{Astronomy Campus, University of Science and Technology, Daedeok-daero 776, Yuseong-gu, Daejeon 34055, Republic of Korea}

\author{Heeseung Zoe}
\email{heeseungzoe@iyte.edu.tr}
\affiliation{Department of Physics, Izmir Institute of Technology, Gulbace, Urla 35430, Izmir, Turkiye}
\affiliation{Department of Physics Education, Pusan National University, Busan 46241, Republic of Korea}
\affiliation{ Department of Physics Engineering, Istanbul Technical University, Maslak 34469, Istanbul, Turkiye}

\begin{abstract}
We consider a simple system consisting of matter, radiation and vacuum components to model the impact of thermal inflation on the evolution of primordial perturbations. The vacuum energy magnifies the primordial modes entering the horizon before its domination, making them potentially observable, and the resulting transfer function reflects the phase changes and energy contents. To determine the transfer function, we follow the curvature perturbation from well outside the horizon during radiation domination to well outside the horizon during vacuum domination and evaluate it on a constant radiation density hypersurface, as is appropriate for the case of thermal inflation. The shape of the transfer function is determined by the ratio of vacuum energy to radiation at matter-radiation equality, which we denote by $\upsilon$, and has two characteristic scales, $k_{\rm a}$ and $k_{\rm b}$, corresponding to the horizon sizes at matter radiation equality and the beginning of the inflation, respectively. 
If $\upsilon \ll 1$, the universe experiences radiation, matter and vacuum domination eras and the transfer function is flat for $k \ll k_{\rm b}$,
 oscillates with amplitude $1/5$ for $ k_{\rm b} \ll k \ll k_{\rm a}$ 
 and oscillates with amplitude $1$ for $k \gg k_{\rm a}$. 
 For $\upsilon \gg 1$, the matter domination era disappears, and the transfer function reduces to being flat for $k \ll k_{\rm b}$ and oscillating with amplitude $1$ for $k \gg k_{\rm b}$.  
\end{abstract}

\maketitle

\newpage 

\section{Introduction}

Inflation provides a theoretical ground for our understanding of the universe \cite{Gliner:1966, Gliner:1970, Guth:1980zm, Linde:1981mu, Albrecht:1982wi}. It makes the universe homogeneous, isotropic, and flat and dilutes unwanted or unobserved relics such as monopoles. It has been constrained by observations of the large-scale structure (LSS) and cosmic microwave background (CMB) \cite{Akrami:2018odb}. 
In supersymmetric cosmology, however, the moduli fields are dangerous to the big bang nucleosynthesis (BBN), if not effectively removed \cite{Coughlan:1983ci, Banks:1993en, deCarlos:1993jw}.  
There have been attempts to solve the moduli problem by arranging low-energy inflation after the primordial inflation \cite{Randall:1994fr}, but it is tricky to control the moduli density in the permissible range  
because the moduli are regenerated after the low-energy inflation. 

This moduli problem can be solved by introducing a {\it thermal inflation} \cite{ 
Lyth:1995hj, Lyth:1995ka}, a brief and secondary inflationary phase after the primordial inflation, being realized by thermal effects on flat directions in supersymmetric theories \cite{Yamamoto:1985mb, Yamamoto:1985rd, Enqvist:1985kz, Bertolami:1987xb, Ellis:1986nn, Ellis:1989ii}. 
The thermal inflation occurs at the {\it primordial dark period} between the end of primordial inflation and BBN. 
Some mechanisms, including reheating/preheating and baryogenesis/leptogenesis, have been suggested for this period.
Thermal inflation scenario has a very different post-inflationary history from the standard scenario and provides new predictions about the primordial dark period.  

While there are some prospects from gravitational waves \cite{Boyle:2005se, Giovannini:2008tm,  Kuroyanagi:2011fy} and collider physics  \cite{Chun:2017spz}, it is hard to directly probe these post-inflationary mechanisms in ``both'' standard and thermal inflation scenarios by observations up to now. 
Thermal inflation, however, gives us a better chance:  {\it 
it magnifies modes that have entered the horizon during the primordial dark period and becomes sensitive to the physics there.} 
In \cite{Hong:2015oqa}, we studied these effects at scales smaller than the horizon size at the beginning of thermal inflation.  
Thermal inflation suppresses the power spectrum of those modes and hence gives
the suppression of CMB $\mu$-distortions \cite{Cho:2017zkj}, the 21-cm hydrogen power spectrum at or before the epoch of reionization, and the formation of galaxy substructures \cite{Hong:2017knn}.

In this paper, we consider a system consisting of matter, radiation and vacuum components to model the impact of thermal inflation on the evolution of primordial perturbations. By assuming that the primordial inflation generates cosmological perturbations in a standard way, we study the growth of the perturbations affected by the secondary thermal inflation. 
We demonstrate how the cosmological scenario between the primordial and thermal inflations affects the power spectrum and possibly leaves observable small-scale features. 
We consider the ratio of vacuum energy to radiation at matter-radiation equality as a key parameter for characterizing the density perturbation, which explains the previous result of \cite{Hong:2015oqa}. 
The detail of moduli dynamics during its domination between the two inflations also impacts the power spectrum, but it is beyond the scope of the present paper.

This paper is organized as follows. 
In section \ref{sec:ti}, we review the thermal inflation scenario and introduce the setting for the perturbation analysis.
In section \ref{sec:TR}, we examine a simple system of vacuum energy and radiation to model thermal inflation.  In section \ref{sec:TRM}, we study a system of vacuum energy, radiation, and matter to calculate the curvature perturbations for thermal inflation analytically. 
In section \ref{sec:transfer}, we present numerical calculations of the power spectrum according to the initial ratio between matter and radiation. 
In section \ref{sec:dis}, we summarize the results and discuss the future work.

\section{Review of Thermal Inflation}\label{sec:ti}

\begin{figure}[bt] 
\begin{center}
\includegraphics[width=0.9\linewidth]{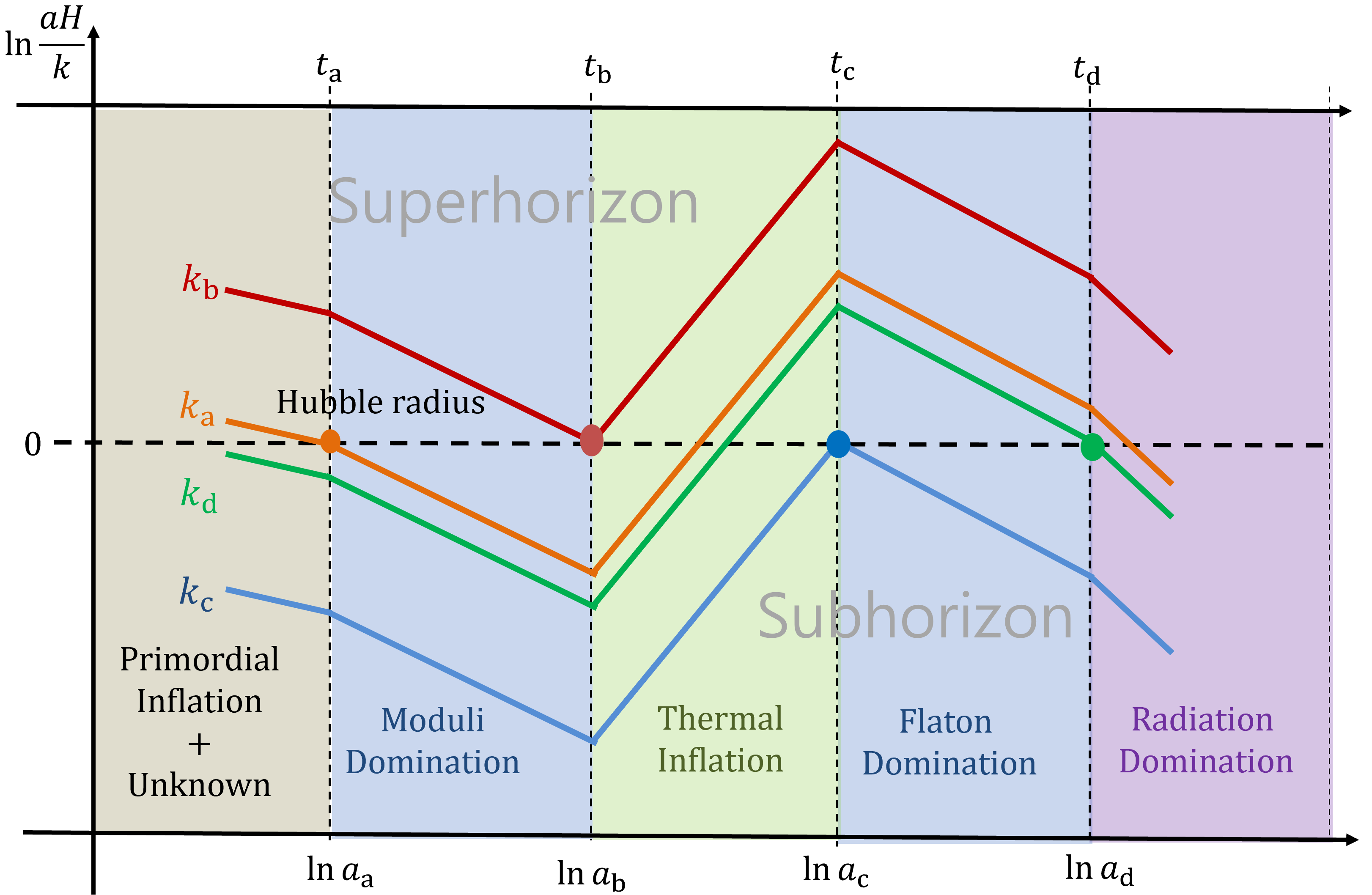}
\end{center}
\caption{The cosmological history of thermal inflation and four characteristic scales.}
\label{fig:history}
\end{figure}

In this section, we briefly review the thermal inflation scenario \cite{Lyth:1995hj,Lyth:1995ka} and introduce density perturbations as a setting for the next sections (for the details, \cite{Hong:2015oqa}). \Fref{fig:history} shows the cosmological history with thermal inflation, and the universe experiences the following four major phases. 

\begin{itemize}

\item {\bf Primordial inflation}: At $t< t_{\rm a}$, the primordial inflation generates the scale-invariant power spectrum, and unknown post-inflationary era including the reheating process follows. 

\item {\bf Moduli domination}: In supersymmetric theories,
the moduli mass in vacuum is an order of the soft supersymmetry breaking scale, i.e.,  $m_\mathrm{mod} \sim m_\mathrm{s} \sim 10^3~\mathrm{to}~10^4\GeV$. At $t \sim t_\mathrm{a}$, the Hubble scale gets $H \sim m_\mathrm{s}$, and the moduli start dominating over the universe.
At $t > t_\mathrm{a}$ or $H < m_\mathrm{s}$, the moduli start oscillating with Planckian amplitude, and the oscillation may be long-lived to spoil the BBN.

\item {\bf Thermal inflation}: At $t = t_{\rm b}$, thermal inflation begins to resolve the moduli problem.
Thermal inflation is realized by flaton(s) that is trapped by finite temperature potential induced by radiation in the universe. The flaton potential is $\fn{V}{\phi,T} = V_0 + \frac{1}{2} \left( \sigma^2 T^2 - m_\phi^2 \phi^2  \right) + \cdots$ where the thermal coupling $\sigma$ is not small and the flaton mass is $m_\phi \sim m_\mathrm{s}$. 
For $V^{1/4}_0 \sim 10^5~\mathrm{to}~10^8\GeV$, thermal inflation has e-folds $N_{\rm bc} \sim 10~\mathrm{to}~15$ and dilute the pre-existing moduli by $\Delta_{\rm TI} \sim e^{3 N_{\rm bc}}$. 

\item {\bf Flaton domination}: As the temperature of the universe drops,
thermal inflation ends until $t \sim t_\mathrm{c}$, and the flaton starts oscillation around its vacuum expectation value $\phi_{\rm vev} \sim \sqrt{V_0}/m_\phi$ and dominates over the universe as a matter phase at $t \gtrsim t_\mathrm{c}$. The moduli can be regenerated after thermal inflation but are diluted further by $\Delta_{\rm flaton} \sim e^{N_{\rm cd}}$ due to the flaton decay. 

\item {\bf Radiation domination}: 
At $t \sim t_\mathrm{d}$, flaton decays to yield the standard radiation dominated universe,
at the temperature $T_\mathrm{d} \sim 10^{-2}~\mathrm{to}~10^2\GeV$.
The universe steps into the standard cosmic history after these radiation domination are recovered.

\end{itemize}
In summary, three distinct extra eras of thermal inflation - moduli domination, thermal inflation, and flaton domination - are inserted between the primordial inflation and radiation domination of the standard scenario. Therefore, the cosmological perturbation evolves differently from the standard inflation scenario. 

We introduce four characteristic scales, $k_\mathrm{a}$, $k_\mathrm{b}$, $k_\mathrm{c}$ and $k_\mathrm{d}$,
where
\begin{equation} \label{kx}
k_x \equiv a_x H_x
\end{equation}
corresponds to the comoving scale of the horizon at the era boundary $t_x$. 
In \Fref{fig:history}, modes with $ k < k_\mathrm{b}$ remain outside the horizon throughout the thermal inflation eras and are not affected by thermal inflation. Modes with $k_\mathrm{b} < k < k_\mathrm{a} $ enter the horizon during moduli domination, and their growth is modified.
Modes with $k > k_\mathrm{a}$ enter the horizon before moduli domination and so {\it probe} that unknown era. Modes with $k > k_\mathrm{d}$ reenter the horizon during flaton domination and so will be twice modified.
Modes with $k > k_{\rm c}$ never exit the horizon throughout thermal inflation eras.

Now we review the perturbation analysis in \cite{Hong:2015oqa}. For $t \lesssim t_{\rm a}$, the post-inflationary physics of the reheating and modulogenesis before the moduli domination is unknown. 
However, for 
\begin{align}
& t_\mathrm{a} \ll t < t_\mathrm{c}~, \label{t} \\
& k \ll k_\mathrm{a}~, \label{kcond}
\end{align}
we can study the perturbations based on a simple system of moduli matter (m), thermal radiation (r) and vacuum energy ($V_0$). The energy density and pressure are 
\begin{align}
&\rho = \rho_\mathrm{m} + \rho_\mathrm{r} + V_0
\label{rho} \\
& p = \frac{1}{3} \rho_\mathrm{r} - V_0~.
\label{p}
\end{align}
The scalar part of the metric perturbation is \cite{Kodama:1984ziu}
\begin{equation}
\tilde{ds}^2 = (1+2A) \d{t}^2 - 2 B_{,i} \d{t} \d{x^i} - \left[ (1+2\mathcal{R}) \fn{a^2}{t} \delta_{ij} + 2 C_{,ij} \right] \d{x^i} \d{x^j} ~.
\end{equation}
We describe the perturbations in moduli and radiation by introducing the gauge invariant variables
\begin{align}
&\Rm  \equiv  \mathcal{R} - \frac{H}{\dot\rho_\mathrm{m}} \delta\rho_\mathrm{m} \\
&\Rr  \equiv  \mathcal{R} - \frac{H}{\dot\rho_\mathrm{r}} \delta\rho_\mathrm{r} 
\end{align}
where $\Rm$ is the curvature perturbation on constant moduli density hypersurfaces and $\Rr$ is the curvature perturbation on constant radiation density hypersurfaces. 
We get two coupled equations for scalar perturbations \cite{Hong:2015oqa}
\begin{align} \label{rrhom} &
\ddot{\mathcal{R}}_{\delta\rho_\mathrm{m}} + H \left( 2 + \frac{\rho_\mathrm{m}}{\rho+p+\frac{2}{3}q^2} \right) \dot{\mathcal{R}}_{\delta\rho_\mathrm{m}}
- \frac{1}{3} q^2 \left( \frac{\rho_\mathrm{m}}{\rho+p+\frac{2}{3}q^2} \right) \Rm
\nonumber \\ & {}
= - \frac{\frac{4}{3} \rho_\mathrm{r}}{\rho+p+\frac{2}{3}q^2} \left( H \dot{\mathcal{R}}_{\delta\rho_\mathrm{r}} - \frac{1}{3} q^2 \Rr \right) ,
\\[1ex] \label{rrhor} &
\ddot{\mathcal{R}}_{\delta\rho_\mathrm{r}} + H \left( 1 + \frac{\frac{8}{3} \rho_\mathrm{r}}{\rho+p+\frac{2}{3}q^2} \right) \dot{\mathcal{R}}_{\delta\rho_\mathrm{r}}
+ \frac{1}{3} q^2 \left( 1 - \frac{\frac{8}{3} \rho_\mathrm{r}}{\rho+p+\frac{2}{3}q^2} \right) \Rr
\nonumber \\ & {}
= - \frac{2\rho_\mathrm{m}}{\rho+p+\frac{2}{3}q^2} \left( H \dot{\mathcal{R}}_{\delta\rho_\mathrm{m}} - \frac{1}{3} q^2 \Rm \right) ,
\end{align}
where $q \equiv k/a$. 
For $t_\mathrm{a} \ll t$, we have
\begin{equation}
\rho_\mathrm{r} \ll \rho_\mathrm{m}\label{mdomr}
\end{equation}
and simplify \eqs{rrhom}{rrhor} to
\begin{align}
\ddot{\mathcal{R}}_{\delta\rho_\mathrm{m}}
+ H \left( 2 + \frac{\rho_\mathrm{m}}{\rho_\mathrm{m} + \frac{2}{3} q^2} \right)\dot{\mathcal{R}}_{\delta\rho_\mathrm{m}}
- \frac{1}{3} q^2 \left( \frac{\rho_\mathrm{m}}{\rho_\mathrm{m} + \frac{2}{3} q^2} \right) \Rm = 0 
\label{oldRm}
\\
\ddot{\mathcal{R}}_{\delta\rho_\mathrm{r}} + H \dot{\mathcal{R}}_{\delta\rho_\mathrm{r}} + \frac{1}{3} q^2 \Rr = F
\label{oldRr}
\end{align}
with
\begin{equation}
F = - 2 \left( \frac{\rho_\mathrm{m}}{\rho_\mathrm{m} + \frac{2}{3} q^2} \right) \left( H \dot{\mathcal{R}}_{\delta\rho_\mathrm{m}} - \frac{1}{3} q^2 \Rm \right).
\end{equation}
With the adiabatic condition 
\begin{align}
\Rm &= \Rr ,
\\
{\dot{\mathcal{R}}}_{\delta \rho_{\rm m}} &= {\dot{\mathcal{R}}}_{\delta \rho_{\rm r}} =0 
\end{align} 
we solve \eqs{oldRm}{oldRr} analytically.

The phase transition of thermal inflation is controlled by the temperature of the radiation of the universe.  Hence, the curvature perturbation at the end of thermal inflation is equal to $\Rr$ and gives the power spectrum by
\begin{equation}
\fn{P}{k} = \fn{P_{\rm pri}}{k} \fn{\mathcal{T}^2}{\frac{k}{k_{\rm b}}}
\end{equation}
where  $\fn{P_{\rm pri}}{k}$ is the power spectrum of the primordial inflation and
\begin{multline}\label{transfer}
\fn{\mathcal{T}}{\frac{k}{k_{\rm b}}} = \cos \left[ \left( \frac{k}{k_{\rm b}} \right) \int_0^\infty \frac{\d\alpha}{\sqrt{\alpha(2+\alpha^3)}} \right] \\
+ 6 \left( \frac{k}{k_{\rm b}} \right) \int_0^\infty \frac{\d\gamma}{\gamma^3} \int_0^\gamma \d\beta \left( \frac{\beta}{2+\beta^3}\right)^{3/2}
\sin \left[ \left( \frac{k}{k_{\rm b}} \right) \int_\gamma^\infty \frac{\d\alpha}{\sqrt{\alpha(2+\alpha^3)}} \right]~.
\end{multline}
is the transfer function summarizing the effects of  thermal inflation eras on the evolution of perturbations.

In this paper, we extend the analysis of \cite{Hong:2015oqa} to $t \lesssim t_{\rm a}$ and $ k \gtrsim k_{\rm a}$ (in contrast to \eqs{t}{kcond}) by treating the moduli as simple matter, $\rho_m \propto a^{-3}$. 
We study \eqs{rrhom}{rrhor} in the limit $k \gg k_{\rm a}$ in section \ref{sec:TR} and  \ref{sec:TRM}.
We find {\it numerical} solutions for \eqs{rrhom}{rrhor} in section \ref{sec:transfer} (in comparison with the {\it analytic} solution of \eq{transfer} for \eqs{oldRm}{oldRr} in \cite{Hong:2015oqa}).

\section{Perturbation for thermal inflation plus radiation}\label{sec:TR}

To model thermal inflation, we first consider the minimal system of radiation and vacuum
\begin{equation}
\rho = \rho_{\rm r} + V_0 \, .
\end{equation}
The curvature perturbation on uniform radiation density hypersurfaces satisfies
\begin{eqnarray}
{\ddot{\mathcal{R}}}_{\delta{\rho}_\mathrm{r}}
+ H \left( 1 + \frac{4\rho_{\rm r}}{2\rho_{\rm r}+q^2} \right){\dot{\mathcal{R}}}_{\delta{\rho}_\mathrm{r}}
+ \frac{1}{3} q^2 \left( 1 - \frac{4\rho_{\rm r} }{2\rho_{\rm r}+q^2} \right) \Rr
\simeq 0
\label{rrhor_bigUpsilon}
\end{eqnarray}
from \eq{rrhor} and we take a growing mode initial condition
\begin{equation}
\Rr =  \mathcal{R}_0 \comma 
{\dot{\mathcal{R}}}_{\delta \rho_{\rm r}} = 0
\label{TR:initial}
\end{equation}
well outside the horizon during radiation domination. 

We define the moment  
when thermal inflation begins, $t_{\rm b}$, by 
\begin{equation}
\fn{\ddot{a}}{t_{\rm b}} \equiv 0~,
\end{equation}
or,
\begin{equation}
\fn{\rho_{\rm r}}{t_{\rm b}} = V_0~.
\end{equation}
From the characteristic scale at $t = t_{\rm b}$
\begin{equation}
k_{\rm b} =  a_{\rm b} H_{\rm b} \label{eq:kb}
\end{equation}
we introduce new parameters
\begin{equation}
\beta \equiv \frac{a}{a_{\rm b}}
\end{equation}
\begin{equation}
\lambda \equiv  \frac{k}{k_{\rm b}}~,
\end{equation}
the energy density is parametrized by
\begin{equation}
\fn{\rho}{\beta} = V_0 \left( 1 + \beta^{-4}  \right)~,
\end{equation}
and \eq{rrhor_bigUpsilon} becomes
\begin{equation}
\label{eq:RrhorA=1}
\beta^2\frac{d^2\Rr}{d\beta^2} +  2\left( 1- \frac{1}{1 + \beta^4}  +   \frac{ 1}{1 +  \frac{\lambda^2}{3}\beta^2}  \right) \beta \frac{d\Rr}{d\beta}
+ \frac{2 \lambda^2\beta^2 }{ 3 \left( 1 + \beta^4 \right) } \left( 1- \frac{ 2 }{1 + \frac{\lambda^2}{3}\beta^2 } \right) \Rr =0~.
\end{equation}
Now we study \eq{eq:RrhorA=1} on large scale $\lambda \ll 1$ ($k \ll k_{\rm b}$) and on small scales $\lambda \gg 1$ ($k \gg k_{\rm b}$) as follows.

\subsection{Large scales $\lambda \ll 1$}

For modes that remain outside the horizon, \eq{eq:RrhorA=1} reduces to 
\begin{equation}
\beta^2 \frac{d^2\Rr}{d\beta^2} 
+2 \left( 2- \frac{1}{1 + \beta^4}  \right) \beta \frac{d\Rr}{d\beta}
\simeq \frac{2 \lambda^2 \beta^2 }{ 3 } 
\left( \beta \frac{d\Rr}{d\beta}  + \frac{1}{1+\beta^4}\Rr \right)~,
\end{equation}
which has a solution
\begin{equation}
\Rr = 
\mathcal{R}_0 \left[ 1 + \frac{2}{3}\lambda^2 \int_0^\beta \frac{ \xi^2d\xi}{\sqrt{1+\xi^4}} \int_{\xi}^\beta \frac{d\eta}{\eta^2\sqrt{1+\eta^4}} 
+\fn{\mathcal{O}}{\lambda^4}
\right]~.
\label{TR:final}
\end{equation}

\subsection{Small scales $\lambda \gg 1$}

For modes that enter the horizon during radiation domination, we solve \eq{eq:RrhorA=1} in two overlapping regimes:

\begin{description}

\item [Radiation domination]

For $\beta \ll 1$, the modes start well outside the horizon during radiation domination and end well inside the horizon during radiation domination.\eq{eq:RrhorA=1} reduces to
\begin{equation}
\beta^2\frac{d^2\Rr}{d\beta^2} +\left( \frac{2}{1+\frac{1}{3}\lambda^2 \beta^2}  \right)\beta \frac{d\Rr}{d\beta}
+ \frac{2}{3}\lambda^2 \beta^2
\left(1-\frac{2}{1+\frac{1}{3}\lambda^2 \beta^2}  \right) \Rr =0~,
\label{TR:Rr0eq}  
\end{equation}
which has a solution
\begin{multline}
\Rr = A_1 \left[ \frac{\sqrt{6}}{\lambda \beta} \sin  \left( \sqrt{\frac{2}{3}}\lambda \beta \right)
- \cos\left( \sqrt{\frac{2}{3}}\lambda \beta \right) \right]  \\
+ B_1 \left[ \frac{\sqrt{6}}{\lambda \beta} \cos \left( \sqrt{\frac{2}{3}}\lambda \beta \right)
+  \sin \left( \sqrt{\frac{2}{3}}\lambda \beta \right) \right]~.
\label{TR:Rr0}
\end{multline}

Matching to the initial condition of \eq{TR:initial} at $\beta \ll \lambda^{-1}$ gives
\begin{equation}
A_1 = \mathcal{R}_0 \comma B_1 = 0~. 
\end{equation}

\item [Well inside the horizon during radiation domination to vacuum domination] For $\beta \gg \lambda^{-1}$, the modes start well inside the horizon during radiation domination and end well outside the horizon during vacuum domination. 
\eq{eq:RrhorA=1} reduces to
\begin{equation}
\left(1 + \beta^4\right)\frac{d^2\Rr}{d\beta^2} +  2\beta^3 \frac{d\Rr}{d\beta}
+ \frac{2 \lambda^2 }{ 3 }  \Rr =0~,
\end{equation}
which has a solution
\begin{align}
\Rr =  A_2 \cos\left[ \lambda \sqrt{\frac{2}{3}} \int_0^{\beta}\frac{1}{\sqrt{1+\xi^4}} d\xi \right] + B_2 \sin \left[ \lambda \sqrt{\frac{2}{3}} \int_0^{\beta}\frac{1}{\sqrt{1+\xi^4}} d\xi \right]~.\label{TR:Rr1}
\end{align}

Matching to \eq{TR:Rr0} at $\lambda^{-1} \ll \beta \ll 1$ gives 
\begin{equation}
A_2 = - A_1 = -\mathcal{R}_0 \comma B_2 = B_1 = 0~.
\end{equation}

\end{description}

\section{Perturbations for thermal inflation plus 
radiation and matter}\label{sec:TRM}

To model the primordial thermal bath, the moduli generated thereby, and thermal inflation, we consider a system of radiation, matter and vacuum 
\begin{equation}
\rho = \rho_{\rm r} + \rho_{\rm m} + V_0 = \rho_0 \left( \frac{a_{\rm a}}{a}\right)^4 + \rho_0 \left( \frac{a_{\rm a}}{a}\right)^3 + V_0~,
\label{eq:rho}
\end{equation}
where subscript ``${\rm a}$'' indicates the moment when the matter energy density surpasses the radiation energy density. 
We describe the changes of three phases in the thermal inflation scenario by \eq{eq:rho}. 
The primordial inflation is followed by radiation-dominated phase, and then moduli dominates over the universe leaving a matter phase. Thermal inflation occurs to dilute the moduli matter and drives a vacuum domination. 
However, \eq{eq:rho} could lead to two phases --- radiation and vacuum phases for $V_0/\rho_o \gg 1$ as follows.  

\begin{figure}[bt] 
\begin{center}
\includegraphics[width=0.8\linewidth]{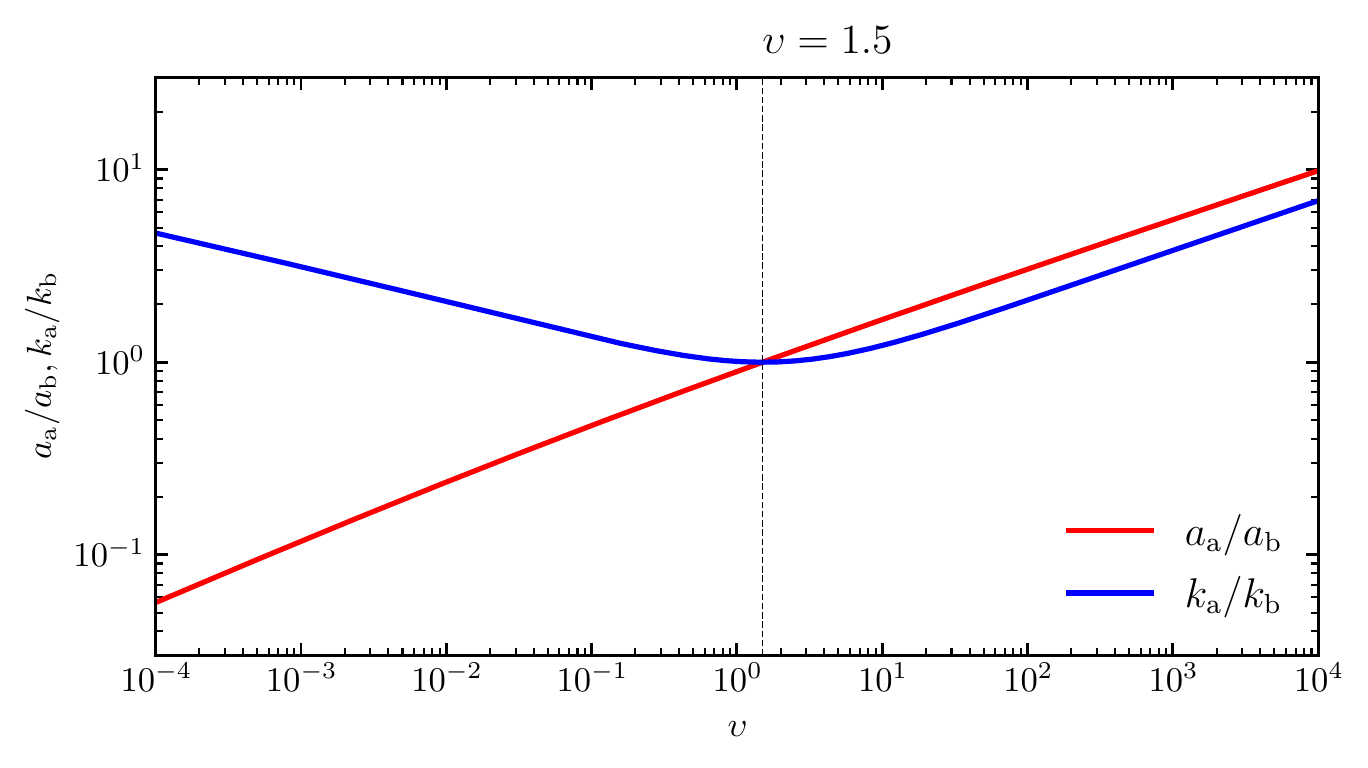}
\end{center}
\caption{The ratios $k_{\rm a}/k_{\rm b}$ and $a_{\rm a}/a_{\rm b}$ as a function of $\upsilon$.}
\label{fig:up_a_k}
\end{figure}

We introduce a useful parameter
\begin{equation}\label{eq:gamma1}
\upsilon \equiv \frac{V_0}{\rho_0}~
\end{equation}
which is the ratio of vacuum energy to radiation at matter-radiation equality.   
The moment 
when thermal inflation begins, $t_{\rm b}$, is found by 
\begin{equation}
\fn{\ddot{a}}{t_{\rm b}} \equiv 0~,
\end{equation}
giving 
\begin{equation}\label{eq:alphab}
 2 \left( \frac{a_{\rm a}}{a_{\rm b}}\right)^4  +  \left( \frac{a_{\rm a}}{a_{\rm b}}\right)^3 - 2 \upsilon = 0
\end{equation}
and
\begin{equation}\label{eq:kappab}
\left( 2 + \upsilon \right) \left( \frac{k_{\rm b}}{k_{\rm a}} \right)^2 = \left( \frac{a_{\rm b}}{a_{\rm a}}\right)^2 \left[ \left( \frac{a_{\rm a}}{a_{\rm b}}\right)^{4}
+\left( \frac{a_{\rm a}}{a_{\rm b}}\right)^{3}
+ \upsilon \right]~,
\end{equation}
where 
\begin{equation}
k_{\rm a} = a_{\rm a} H_{\rm a}
\end{equation}
and $k_{\rm b} = a_{\rm b} H_{\rm b}$ is already given in \eq{eq:kb}. 
From \eqs{eq:alphab}{eq:kappab}, the relations between $\upsilon$, $a_{\rm a}/a_{\rm b}$ and $k_{\rm a}/k_{\rm b}$ are fixed by the value of $\upsilon$ in \Fref{fig:up_a_k}.

\begin{figure}[bt] 
\begin{center}
\includegraphics[width=0.8\linewidth]{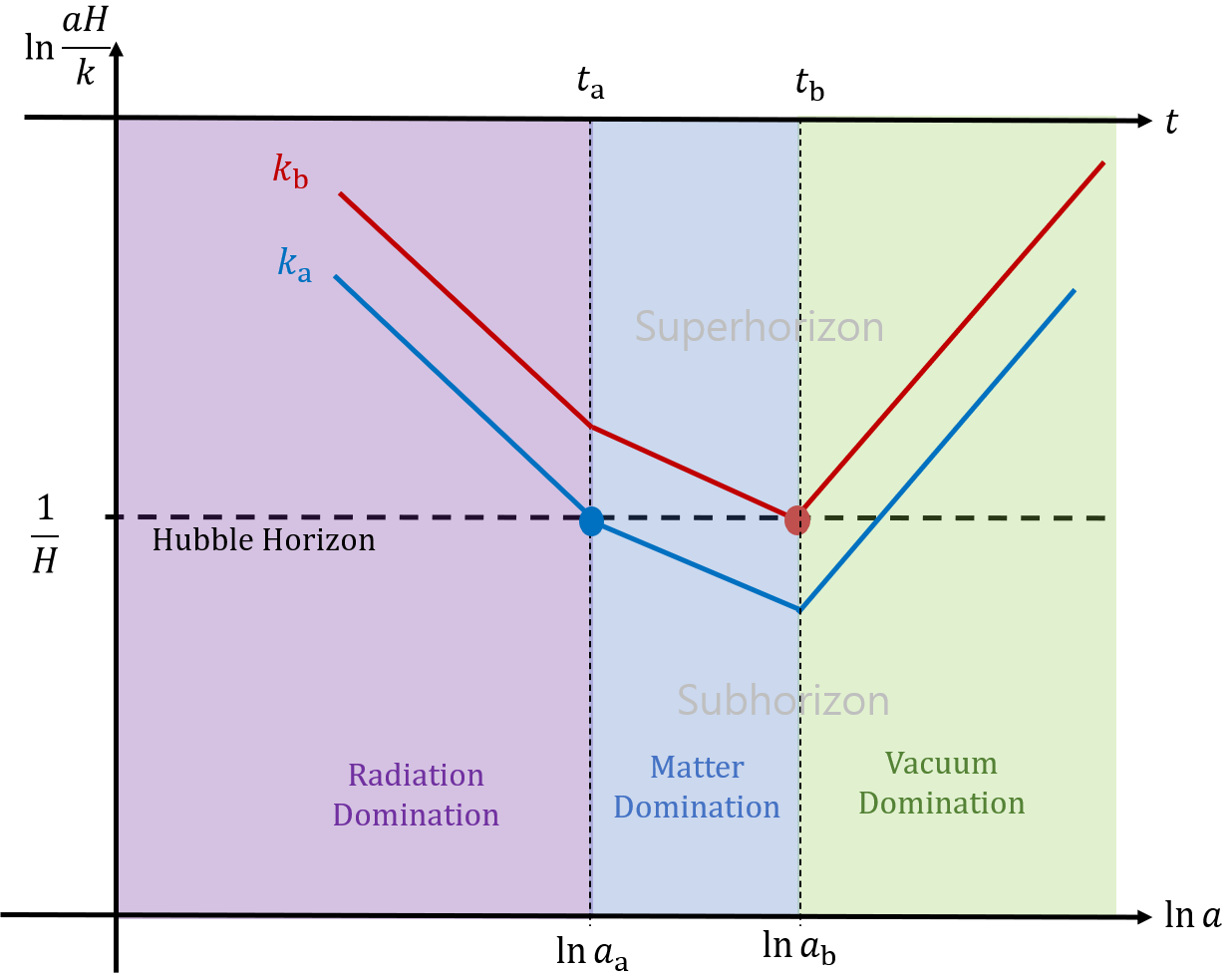}
\end{center}
\caption{ A diagram showing the evolution of horizon scales for $\upsilon \ll 1$. 
The characteristic scales $k_{\rm a}$ and $k_{\rm b}$ are shown with the phase evolution. }
\label{fig:3steps}
\end{figure}

For $\upsilon \ll 1$, \eqs{eq:alphab}{eq:kappab} produces asymptotic formulae
\begin{equation}
\frac{a_{\rm a}}{a_{\rm b}}  \simeq \left( 2\upsilon \right)^{\frac{1}{3}} 
\comma 
\frac{k_{\rm a}}{k_{\rm b}} \simeq  \left(\frac{2^5}{3^3\upsilon}\right)^\frac{1}{6}~, 
\label{eq:kakbupgg1}
\end{equation}
implying $a_{\rm a} \ll a_{\rm b}$ and $k_{\rm b} \ll k_{\rm a}$. 
In this case, we have a radiation phase initially at $a \ll a_{\rm a}$, a matter phase at $a_{\rm a} < a < a_{\rm b}$, and then a vacuum phase at $a> a_{\rm b}$ in \Fref{fig:3steps}. These three phases mimic the eras after primordial inflation in thermal inflation scenario. 
In this limit, \eq{eq:rho} gives $\fn{\rho_{\rm r}}{a_{\rm b}} \ll \fn{\rho_{\rm m}}{a_{\rm b}}$, and it reduces to the case of \cite{Hong:2015oqa}.

\begin{figure}[bt] 
\begin{center}
\includegraphics[width=0.8\linewidth]{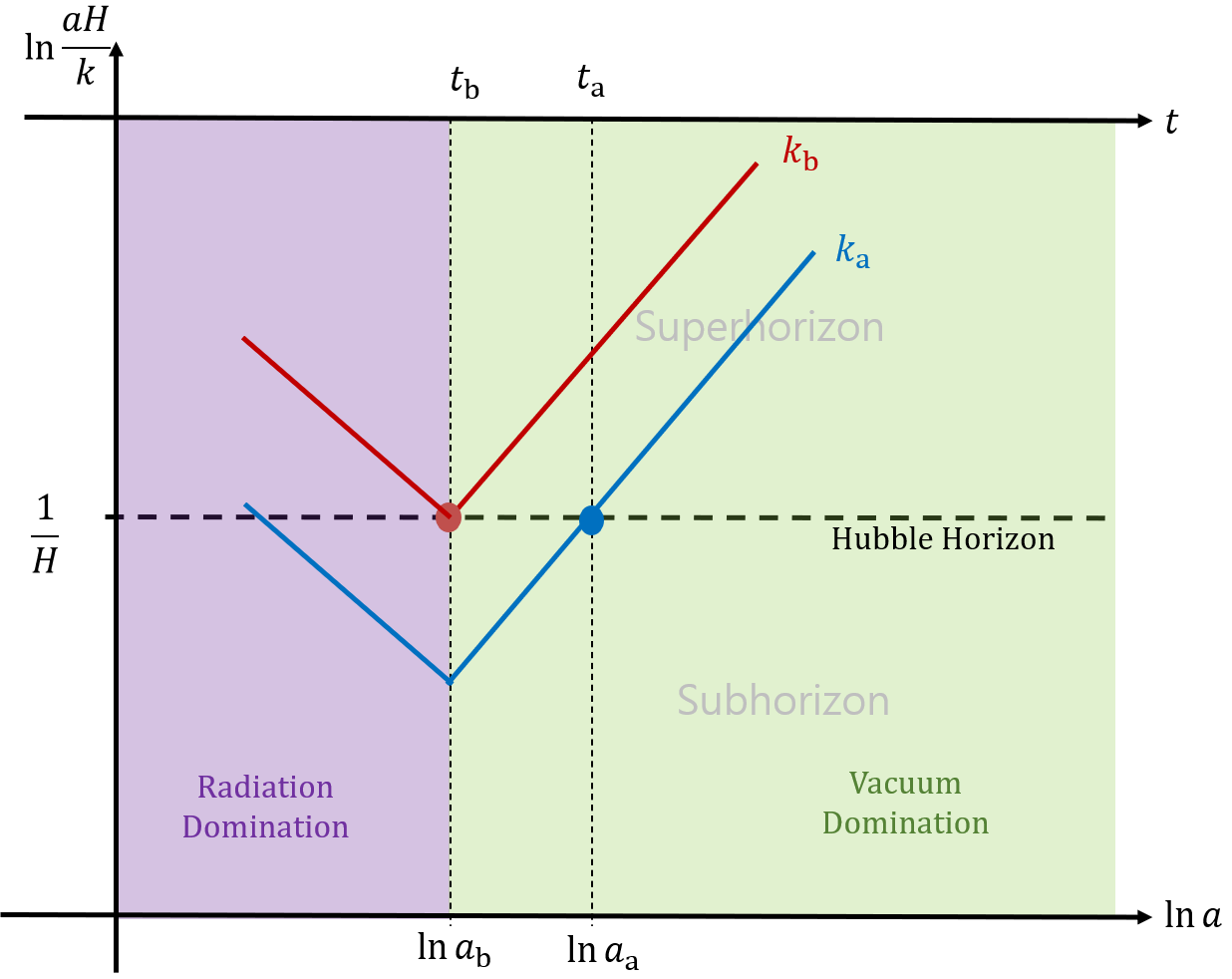}
\end{center}
\caption{A diagram showing the evolution of horizon scales for $\upsilon \gg 1$. 
The characteristic scales $k_{\rm a}$ and $k_{\rm b}$ are shown with the phase evolution. Note that matter-domination phase from \eq{eq:rho} does not exist. }
\label{fig:2steps}
\end{figure}

For $\upsilon \gg 1$, \eqs{eq:alphab}{eq:kappab} produces asymptotic relations 
\begin{equation}
\frac{a_{\rm a}}{a_{\rm b}}  \simeq \upsilon^{\frac{1}{4}} 
\comma 
\frac{k_{\rm a}}{k_{\rm b}} \simeq \left(\frac{\upsilon}{4}\right)^{\frac{1}{4}}~,
\end{equation}
implying $a_{\rm b} \ll a_{\rm a}$ and $k_{\rm b} \ll k_{\rm a}$. 
In this case, we have a radiation phase initially at $a \ll a_{\rm b}$ which turns into a vacuum phase at $ a = a_{\rm b} $. At $a = a_{\rm a}$, however, \eq{eq:rho} gives $\rho_{\rm m} + \rho_{\rm r} \ll V_0$ leaving a vacuum phase.  Hence, \eq{eq:rho} describes only two phases of radiation and vacuum in \Fref{fig:2steps}.  
In this limit, \eq{eq:rho} gives $\fn{\rho_{\rm r}}{a_{\rm b}} \gg \fn{\rho_{\rm m}}{a_{\rm b}}$, and it clearly becomes the case of thermal inflation plus radiation discussed in the previous section.

We take the adiabatic growing mode as the initial condition 
\begin{align}
\Rm &= \Rr = \mathcal{R}_0,\label{adiabatic1}
\\
{\dot{\mathcal{R}}}_{\delta \rho_{\rm m}} &= {\dot{\mathcal{R}}}_{\delta \rho_{\rm r}} =0  \label{adiabatic2}
\end{align} 
well outside the horizon during radiation domination.

We rewrite the equations by introducing new parameters 
\begin{align}
\alpha  &\equiv \frac{a}{a_{\rm a}}~, \\
\kappa &\equiv \frac{k}{k_{\rm a}}~.
\end{align}
Then \eq{rrhor} becomes
\begin{multline}
\alpha^2 \frac{d^2 \Rr}{d\alpha^2}
+ 2 \left( 1 + \frac{1}{1+ \frac{3}{4}\alpha + \frac{2+\upsilon}{6}\kappa^2 \alpha^2} - \frac{1+\frac{3}{4}\alpha}{1+\alpha + \upsilon \alpha^4} \right)  \alpha \frac{d \Rr}{d\alpha}
\\
+ \left( \frac{2+\upsilon}{3}\kappa^2\alpha^2 \right)\left( \frac{1}{1 +\alpha + \upsilon\alpha^4 }  \right) \left(  1- \frac{2}{1 + \frac{3}{4}\alpha + \frac{2+\upsilon}{6}\kappa^2 \alpha^2}  \right)  \Rr
\\
 =  - \frac{3}{2}\left( \frac{\alpha}{1+\frac{3}{4}\alpha +\frac{2+\upsilon}{6}\kappa^2 \alpha^2} \right) \left[ \alpha\frac{d \Rm}{d\alpha} - \left( \frac{2+\upsilon}{3}\kappa^2\alpha^2 \right) \left( \frac{1}{1 + \alpha + \upsilon\alpha^4 } \right)\Rm \right]~,
 \label{rrhor_alpha}
\end{multline}
and \eq{rrhom} becomes 
\begin{multline}
\alpha^2\frac{d^2 \Rm}{d\alpha^2}
+ \left( 3 + \frac{\frac{3}{4}\alpha}{1+\frac{3}{4}\alpha+\frac{2+\upsilon}{6}\kappa^2 \alpha^2} - \frac{2+\frac{3}{2}\alpha}{1 + \alpha + \upsilon \alpha^4} \right)  \alpha \frac{d \Rm}{d\alpha}
\\
- \left( \frac{2+\upsilon}{3}\kappa^2\alpha^2 \right)\left( \frac{1}{1 +\alpha + \upsilon\alpha^4 }  \right) \left(  \frac{\frac{3}{4}\alpha}{1 + \frac{3}{4}\alpha +\frac{2+\upsilon}{6}\kappa^2 \alpha^2}  \right)  \Rm
\\ \label{rrhom_alpha}
 =  - \left( \frac{1}{1+\frac{3}{4}\alpha +\frac{2+\upsilon}{6}\kappa^2 \alpha^2} \right) \left[ \alpha\frac{d \Rr}{d\alpha} - \left( \frac{2+\upsilon}{3}\kappa^2\alpha^2 \right)\left( \frac{1}{1 +\alpha + \upsilon\alpha^4 }  \right)\Rr \right] ~. 
\end{multline}  
Now we study the asymptotic behaviors of \eqs{rrhor_alpha}{rrhom_alpha}  for both $\upsilon \ll 1$ and $\upsilon \gg 1$ on large scale $\kappa \ll 1$ ($k \ll k_{\rm a}$) and on small scales $\kappa \gg 1$ ($k \gg k_{\rm a}$) as follows.

\subsection{Large scales $\kappa \ll 1$}

We consider the cases of $\upsilon \ll 1$ and $\upsilon \gg 1$ separately in the large scale limit.
For
\begin{equation}
\upsilon \ll 1~,
\end{equation} 
and for the late time 
\begin{equation}
\alpha \gg 1 ~,
\end{equation} 
\eqs{rrhor}{rrhom} become 
\begin{multline}
\left( \alpha + \upsilon\alpha^4 \right) \frac{d^2 \Rr}{d\alpha^2} 
+ \frac{1}{2} \left( 1 + 4 \upsilon \alpha^3 \right)\frac{d \Rr}{d \alpha} 
+ \frac{2}{3}\kappa^2 \Rr \\
= - \frac{3}{2} \left(\frac{1}{\frac{3}{4} \alpha +\frac{1}{3}\kappa^2 \alpha^2} \right)
\left[ 
\left( \alpha + \upsilon\alpha^4 \right) \frac{d\Rm}{d\alpha} 
- \frac{2}{3}\kappa^2 \alpha \Rm  \right] \label{Rrsingle}
\end{multline}
\begin{multline}
\alpha^2 \frac{d^2 \Rm}{d\alpha^2} + \left( 3 + \frac{1}{1+\frac{4}{9}\kappa^2 \alpha}- \frac{\frac{3}{2}}{1+\upsilon\alpha^3} \right) \alpha \frac{d \Rm}{d \alpha} \\ 
 - \left( \frac{\frac{2}{3}\kappa^2 \alpha}{ 1 + \upsilon \alpha^3}\right) \left(\frac{1}{1+\frac{4}{9}\kappa^2 \alpha}\right) \Rm = 0 ~,
 \label{Rmsingle}
\end{multline}
which have solutions
\begin{multline}
\Rr  =  A \cos \left[ \sqrt{\frac{2}{3}} \kappa \int_{\alpha_i}^\alpha \frac{d\xi}{\sqrt{ \xi + \upsilon \xi^4}}  \right] 
+ B \sin \left[ \sqrt{\frac{2}{3}} \kappa \int_{\alpha_i}^\alpha \frac{d\xi}{\sqrt{\xi + \upsilon \xi^4}}  \right]  \\
+  \frac{9}{2} \int_{\alpha_i}^\alpha d\xi \frac{\sqrt{\frac{2}{3}}\kappa}{\sqrt{\xi + \upsilon \xi^4} 
} \sin \left[  \sqrt{\frac{2}{3}} \kappa \int_\xi^\alpha \frac{d\sigma}{\sqrt{\sigma + \upsilon \sigma^4}}  \right] \left[ \frac{\fn{\Rm}{\xi} - C }{\kappa^2 \xi} \right] ~,
\label{TRM:smallupsilonRr}
\end{multline}
where $\alpha_i \ll 1$ is the initial value, and
\begin{eqnarray}
\Rm = C \left[ 1 + \frac{2}{3} \kappa^2 \sqrt{\frac{1+\upsilon\alpha^3}{\alpha^3}} \int_0^\alpha \left( \frac{\xi}{1+\upsilon\xi^3}\right)^{\frac{3}{2}} d\xi   \right] 
+ D  \sqrt{\frac{1+ \upsilon \alpha^3}{\alpha^3}}~.
\label{TRM}
\end{eqnarray}

Matching to the initial conditions of \eqs{adiabatic1}{adiabatic2} at $\alpha \ll \kappa^{-2}$ gives
\begin{equation}
A = \mathcal{R}_0 \comma B = 0 \comma C = \mathcal{R}_0 \comma D = 0
\end{equation}
reproducing the result of \cite{Hong:2015oqa}.

For $\upsilon \gg 1$, we return to the case of thermal inflation plus radiation, and \eqs{rrhor}{rrhom} reduce to  \eq{rrhor_bigUpsilon} giving \eq{TR:final} as the proper solution.

\subsection{Small scales $\kappa  \gg 1$}
We study the cases of $\upsilon \ll 1$ and $\upsilon \gg 1$ together in the small scale limit.
For modes that enter the horizon during radiation domination,
we solve \eqs{rrhor_alpha}{rrhom_alpha} in two overlapping regimes: 
\begin{description}
\item [Radiation domination]  For $\alpha \ll \fn{{\rm min}}{1,\upsilon^{-\frac{1}{4}}} $, \eq{rrhor_alpha} becomes
\begin{equation}
\alpha^2 \frac{d^2 \Rr}{d \alpha^2}
+ \left( \frac{2}{1+ \frac{2+\upsilon}{6}\kappa^2 \alpha^2} \right) \alpha \frac{d \Rr}{d \alpha}
+  \frac{2+\upsilon}{3}\kappa^2\alpha^2  \left(  1- \frac{2}{1+ \frac{2+\upsilon}{6}\kappa^2 \alpha^2}  \right)  \Rr  =0 ~,
\label{TRM:Rr0eq} 
\end{equation}
whose solution is given by 
\begin{multline}\label{TRM:Rr0}
\Rr  = 
A_1 \left[ \sqrt{\frac{3}{2+\upsilon}} \left( \frac{2}{\kappa \alpha}\right)
\sin \left( \sqrt{\frac{2+\upsilon}{3}}\kappa \alpha \right)
- \cos \left( \sqrt{\frac{2+\upsilon}{3}}\kappa \alpha \right)  \right] \\
 +   B_1 \left[ \sqrt{\frac{3}{2+\upsilon}} \left( \frac{2}{\kappa \alpha}\right)
\cos \left( \sqrt{\frac{2+\upsilon}{3}}\kappa \alpha \right)
+ \sin \left( \sqrt{\frac{2+\upsilon}{3}}\kappa \alpha \right)  \right] ~.
\end{multline}

Matching to the initial conditions of \eqs{adiabatic1}{adiabatic2} at $\alpha \ll \fn{{\rm min}}{\kappa^{-1}, \kappa^{-1}\upsilon^{-\frac{1}{2}} }$
gives
\begin{equation}
A_1 = \mathcal{R}_0 \comma B_1 = 0 ~.
\end{equation}

\item  [Well inside the horizon during radiation domination to vacuum domination] For $ \alpha \gg \fn{{\rm min}}{\kappa^{-1}, \kappa^{-1}\upsilon^{-\frac{1}{2}} }$, \eq{rrhor_alpha} becomes
\begin{equation}
\left( 1 + \alpha + \upsilon\alpha^4  \right)\frac{d^2\Rr}{d \alpha^2}
+ \frac{1}{2} \left( 1 + 4 \upsilon \alpha^3 \right)\frac{d \Rr}{d \alpha}
+ \left( \frac{2+\upsilon}{3} \right)\kappa^2 \Rr 
= 0 ~,
\label{TRM:Rr1eq}
\end{equation}
whose solution is given by 
\begin{multline}
\Rr = A_2 \cos \left( \sqrt{\frac{2+\upsilon}{3}} \kappa \int_0^{\alpha} \frac{d\xi}{\sqrt{1+\xi + \upsilon \xi^4}}  \right) \\
+  B_2 \sin \left( \sqrt{\frac{2+\upsilon}{3}} \kappa \int_0^{\alpha} \frac{d\xi}{\sqrt{1+\xi + \upsilon \xi^4}}  \right) ~.
\label{TRM:Rr1}
\end{multline}

Matching to \eq{TRM:Rr0} at $  \fn{{\rm min}}{\kappa^{-1}, \kappa^{-1}\upsilon^{-\frac{1}{2}} } \ll \alpha \ll \fn{{\rm min}}{1,\upsilon^{-\frac{1}{4}}} $ gives
\begin{equation}
A_2 = - A_1 = - \mathcal{R}_0 \comma B_2 = B_1 = 0~.
\end{equation}

For $\upsilon \gg 1$, we can show that \eq{TRM:Rr1} becomes \eq{TR:Rr1} by using the relation of \eq{eq:kakbupgg1}.

\end{description}

In the next section, we construct the asymptotic forms of transfer function from the primordial inflationary to thermal inflation power spectrum by using the results of \eqs{TR:final}{TRM:Rr1} for $\upsilon \gg 1$ and \eqs{TRM:smallupsilonRr}{TRM:Rr1} for $\upsilon \ll 1$  to study the numerical solution of \eqs{rrhor_alpha}{rrhom_alpha} for various values of $\upsilon$.

\section{Transfer functions}\label{sec:transfer}

\begin{figure}[bt] 
\begin{center}
\includegraphics[width=\linewidth]{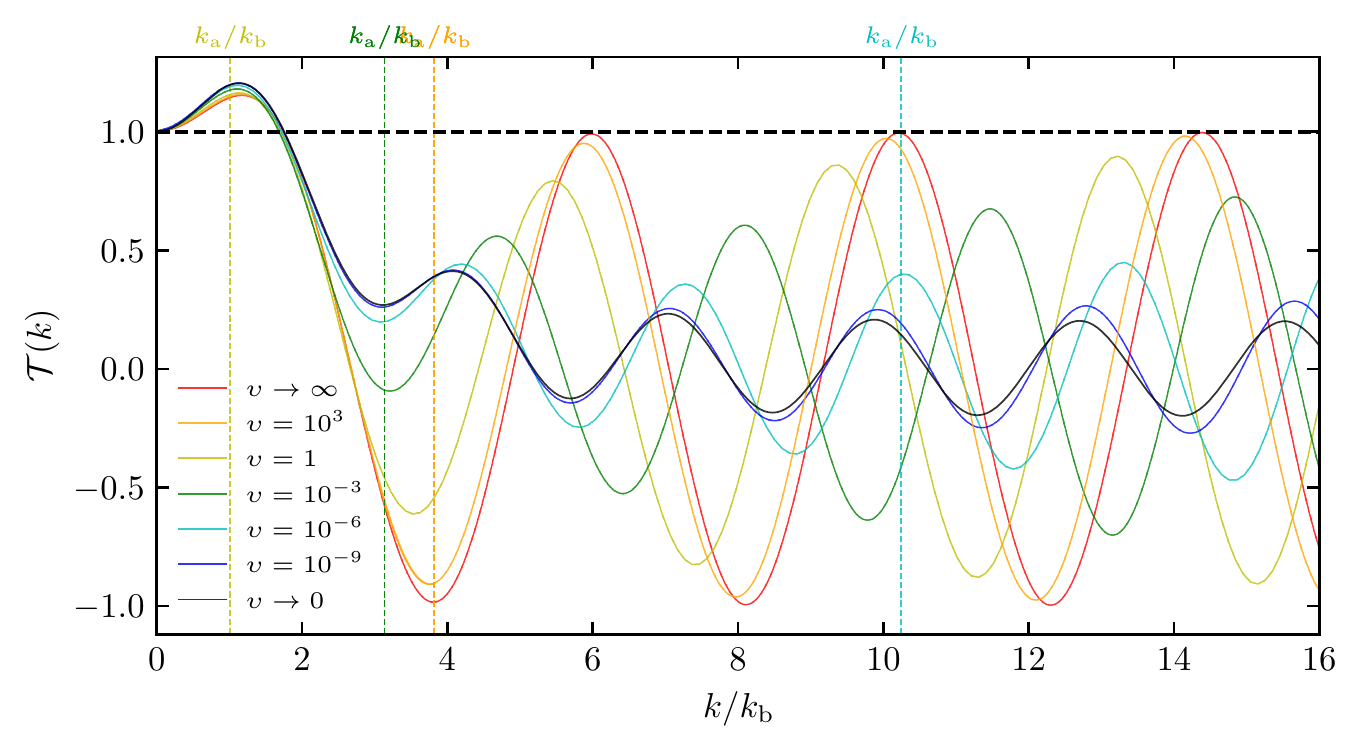}
\end{center}
\caption{Transfer functions of the thermal inflation scenarios with various values of $\upsilon$ as a function of $k/k_{\rm b}$.}\label{fig:transfer_kb}
\end{figure}

We take the adiabatic condition to calculate the density perturbation in \eqs{adiabatic1}{adiabatic2}.
The effects of thermal inflation is summarized by the curvature perturbation on the radiation density hypersurfaces $\Rr$ as    
\begin{equation}
\fn{\mathcal{T}}{k} \equiv \frac{\fn{\Rr}{k, \infty}}{\fn{\Rr}{k, 0}}~.
\end{equation}

For thermal inflation and radiation, we find the asymptotic form of the transfer function from \eqs{TR:final}{TR:Rr1}. On large scale, it is almost scale-invariant and has slight enhancement at $k \simeq k_{\rm b}$,
\begin{equation} \label{TR:kappazerosingle}
\fn{\mathcal{T}}{k} 
\xrightarrow{k \ll k_{\rm b}}
1 + \mu_0 \left( \frac{k}{k_{\rm b}}  \right)^2 ,
\end{equation}
where
\begin{equation}\label{mu0}
\mu_0 = \int_0^\infty \frac{ \xi^2d\xi}{\sqrt{1+\xi^4}} \int_{\xi}^\infty \frac{d\eta}{\eta^2\sqrt{1+\eta^4}} \simeq 0.2393 ~.
\end{equation}
On small scales, it is sinusoidal with an amplitude of unity, 
\begin{equation} \label{TR:kappainftysingle}
\fn{\mathcal{T}}{k} 
\xrightarrow{k \gg k_{\rm b}}
- \cos \left[ \mu_1 \left( \frac{k}{k_{\rm b}} \right) \right]~ ,
\end{equation}
where
\begin{equation}\label{mu1}
\mu_1 = \sqrt{\frac{2}{3}}\int_0^{\infty}\frac{d \beta}{\sqrt{1+\beta^4}} = \frac{\fn{\Gamma}{\frac{1}{4}}^2}{2\sqrt{6\pi}} \simeq 1.5139 ~.
\end{equation}

\begin{figure}[bt] 
\begin{center}
\includegraphics[width=\linewidth]{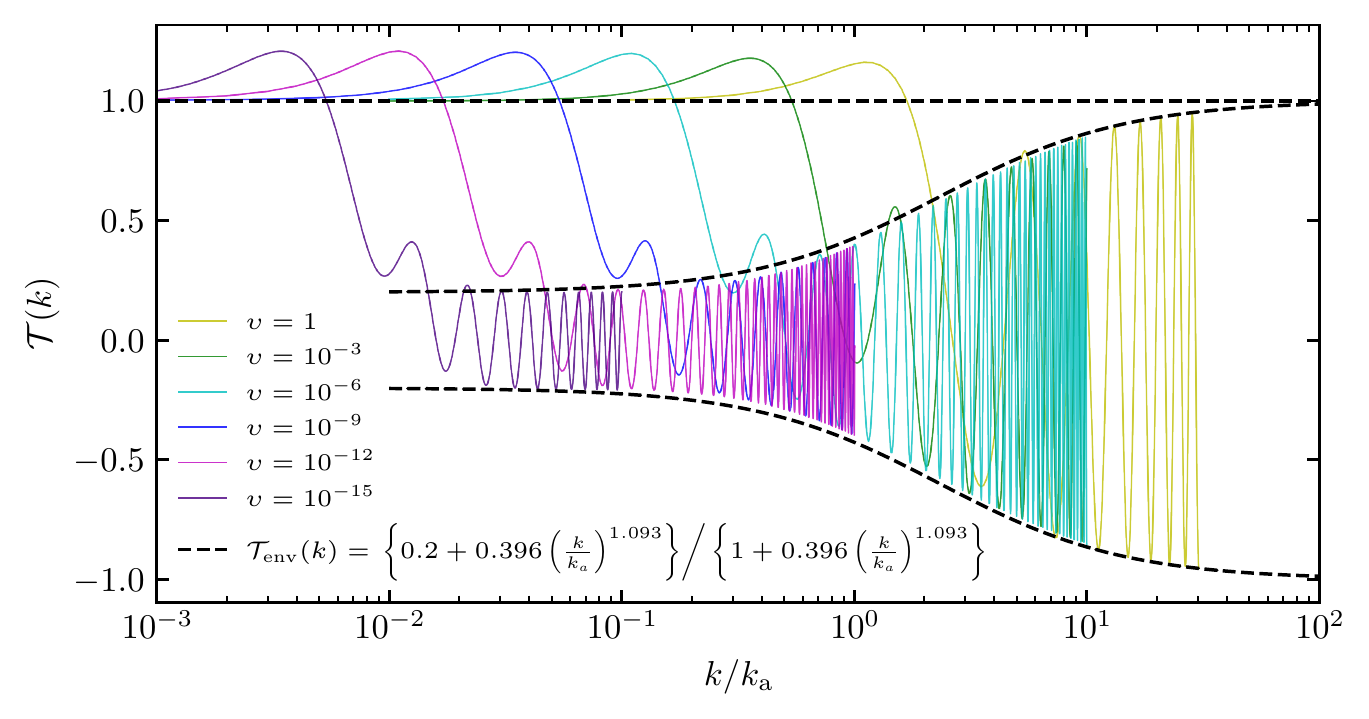}
\end{center}
\caption{Transfer functions of the thermal inflation scenarios with various values of $\upsilon \lesssim 1$ as a function of  $k/k_{\rm a}$.}\label{fig:transfer_ka}
\end{figure}

For thermal inflation, radiation and matter, the transfer function for $\upsilon \ll 1$ becomes 
\begin{multline}
\fn{\mathcal{T}}{k} = 
 \cos \left[\left(\frac{k}{k_{{\rm b}}} \right)   \int_0^\infty \frac{d\xi}{\sqrt{\xi(2+\xi^3)}} \right] \\
 + 6 \left(\frac{k}{k_{{\rm b}}} \right) 
\int_0^\infty \frac{d\eta}{\eta^3} \int_0^\eta d\zeta \left( \frac{\zeta}{2+\zeta^3}\right)^\frac{3}{2} 
\sin \left[ \left(\frac{k}{k_{{\rm b}}} \right)  \int_\eta^\infty \frac{d\xi}{\sqrt{\xi(2+\xi^3)}} \right]
\label{eq:singletransfer}
\end{multline}
from \eq{TRM:smallupsilonRr}, which is exactly the same form of \cite{Hong:2015oqa}. 

On large scales that remain outside the horizon, the transfer function goes to 
\begin{equation}
\fn{\mathcal{T}}{k} \longrightarrow
1 + \nu_0 \left( \frac{k}{k_{\rm b}}\right)^2 ~,
\label{TRM:kappazero}
\end{equation}
where
\begin{equation}
\nu_0 = \int_0^\infty \d\alpha \left( \frac{\alpha}{2+\alpha^3} \right)^\frac{3}{2}
= \frac{2^{7/3}\pi^{3/2}}{3^{3/2} \fn{\Gamma}{\frac{1}{6}} \fn{\Gamma}{\frac{1}{3}}}
\simeq 0.3622 ~.
\end{equation}

On smaller scales that enter the horizon during matter domination, $k_{\rm b} \ll k \ll k_{\rm a}$, the transfer function goes to  
\begin{equation}
\fn{\mathcal{T}}{k} \longrightarrow - \frac{1}{5} \cos \left[\nu_1 \left( \frac{k}{k_{\rm b}}\right) \right] ~,
\label{TRM:kappainfty1}
\end{equation}
where
\begin{equation}
\nu_1 = \int_0^\infty \frac{\d\alpha}{\sqrt{\alpha(2+\alpha^3)}}
= \frac{\fn{\Gamma}{\frac{1}{6}} \fn{\Gamma}{\frac{1}{3}}}{2^{1/3}3\sqrt{\pi}}
\simeq 2.2258 ~.
\end{equation}

On much smaller scales that enter the horizon during radiation domination, $k \gg k_{\rm a}$, \eq{TRM:Rr1} gives the transfer function as
\begin{equation}
\fn{\mathcal{T}}{k}
\longrightarrow
- \cos \left[ \fn{\tau_0}{\upsilon} \sqrt{\frac{2+\upsilon}{3}} \left( \frac{k}{k_{\rm a}} \right)  \right]
\simeq
    \begin{cases}
     -\cos \left[ \nu_1  \left( \frac{k}{k_{\rm b}}  \right) \right] & \text{for $\upsilon \ll 1 $}\\
      -\cos \left[ \mu_1 \left( \frac{k}{k_{\rm b}}  \right) \right] & \text{for $\upsilon \gg 1 $}
    \end{cases}   ~,
    \label{TRM:kappainfty2}    
\end{equation}
where
\begin{equation}
\fn{\tau_0}{\upsilon} = \int_0^{\infty} \frac{d\alpha}{\sqrt{1+\alpha + \upsilon \alpha^4}}
\simeq
    \begin{cases}
      2^\frac{1}{3} \nu_1 \upsilon^{-\frac{1}{6}} -2 + \fn{\mathcal{O}}{\upsilon^\frac{1}{6}}& \text{for $\upsilon \ll 1 $}\\
      \sqrt{\frac{3}{2}}\mu_1 \upsilon^{-\frac{1}{4}} -\frac{1}{4}\upsilon^{-\frac{1}{2}} + \fn{\mathcal{O}}{-\upsilon^\frac{3}{4}} & \text{for $\upsilon \gg 1 $}
    \end{cases}  ~. 
    \label{tau0}    
\end{equation}

The transfer functions are plotted with $k/k_{\rm b}$ in \Fref{fig:transfer_kb}, and they are consistent with the asymptotic forms of \eqsss{TR:kappazerosingle}{TRM:kappazero}{TRM:kappainfty1}{TRM:kappainfty2}.
In \Fref{fig:transfer_ka}, the transfer functions for $\upsilon \lesssim 1$ are plotted with $k/k_{\rm a}$, and  we can find the envelop of the amplitude of transfer functions 
\begin{equation}
\fn{\mathcal{T}_{\rm env}}{k} \simeq \frac{ 0.2 + 0.396 \left( \frac{k}{k_{\rm a}} \right)^{1.093}}{1 + 0.396 \left( \frac{k}{k_{\rm a}} \right)^{1.093}} \longrightarrow
\begin{cases}
0.2, & k_{\rm b} < k \ll k_{\rm a} \\
0.427, & k \simeq k_{\rm a}\\
1, & k_{\rm a} \ll k
\end{cases} ~.
\end{equation}
Note that the oscillating part with the amplitude of $0.2$ can be found only for $\upsilon \ll 1$.
At $k \simeq k_{\rm a}$, the envelop is more precisely approximated by  
\begin{equation} \label{kappaone}
\fn{\mathcal{T}_{\rm env}}{k} \longrightarrow 0.427 + 0.422 \log_{10} \left( \frac{k}{k_{\rm a}} \right)  ~.
\end{equation}

\section{Conclusion}\label{sec:dis}

In this paper, we model density perturbations for thermal inflation by considering a multi-component system of radiation, matter and vacuum energy $\rho = \rho_{\rm r} + \rho_{\rm m} + V_0$. 
The transfer function converting the primordial power spectrum to the model prediction is calculated by tracing the linear evolution of their curvature perturbations in \Fref{fig:transfer_kb} and \Fref{fig:transfer_ka}. 
The value of $ \upsilon 
\equiv V_0 / \rho_0$ is a key parameter governing the shape of transfer functions. $\upsilon$ adjusts the relative ratio between matter and radiation energy densities at the beginning of the second (thermal) inflation.  The system equivalently reduces to the case of \cite{Hong:2015oqa} for $\upsilon \rightarrow 0$ and to the case of vacuum and radiation for $\upsilon \rightarrow \infty$.    

For $\upsilon < 1.5$ (see \Fref{fig:up_a_k}), we considered the following three kinds of modes: 
first, the largest modes with $k < k_{\rm b}$ are always beyond the horizon and their curvature perturbation also remains constant leaving $\mathcal{T} \simeq 1$. 
Second, the intermediate modes with $k_{\rm b} < k < k_{\rm a}$ come into the horizon during moduli matter domination and exit the horizon during thermal inflation. 
For these modes, perturbations are enhanced by $\mathcal{T} \simeq 1.3622$ at $k \simeq k_{\rm b}$ and suppressed by $\left| \mathcal{T} \right| \rightarrow \frac{1}{5} $ on 
smaller scales $k \gg k_{\rm b}$.  
Third, the smallest modes with $k > k_{\rm a}$ enter the horizon during radiation domination and exit during thermal inflation. 
These modes are inside the horizon at radiation domination before the moduli domination and so behave oscillating with the amplitude $\left| \mathcal{T} \right| \rightarrow 1$. 

For $\upsilon  \gg 1$, the matter component is negligible compared to the radiation at the beginning of thermal inflation. The transfer function is constant on large scales $k < k_{\rm b}$, enhanced as $\mathcal{T} \simeq 1.2393$ at $k \simeq k_{\rm b}$, and then have sinusoidal oscillations with an amplitude of unity at $k \gg k_{\rm b}$.

Note that the characteristic scale $k_{\rm b}$ is estimated by 
\begin{equation}
k_{\rm b} \simeq 10^3 \invMpc \left( \frac{e^{20}}{e^{N}}\right) \left( \frac{V_0^\frac{1}{4}}{10^7 \GeV} \right)^\frac{2}{3}
\left( \frac{T_{\rm d}}{\GeV}\right)^\frac{1}{3}
\end{equation}
where $N$ is the e-folds during thermal inflation, 
the vacuum energy for thermal inflation is 
$10^3 \GeV \ll V_0^\frac{1}{4} < 10^{11} \GeV $, 
and the reheating temperature for the radiation domination is $10^{-2} \GeV \lesssim T_{\rm d} \lesssim 10^2 \GeV$ 
in \cite{Hong:2015oqa, Hong:2017knn}, 
and $k_{\rm a} /  k_{\rm b} \lesssim 10$ if 
${10^{-4}} \lesssim \upsilon \lesssim 10^{4}$ 
in \Fref{fig:up_a_k}. Hence, the changes in the transfer functions according to the value of $\upsilon$ could be explored by small-scale observations including CMB spectral distortions  \cite{Cho:2017zkj}, the substructure of galaxies \cite{Hong:2017knn, Leo:2018kxp,Enqvist:2019jkb}, and the 21-cm hydrogen line \cite{Hong:2017knn, Leo:2019gwh}.
Our results can be applied to the calculation for the density perturbations in multiple inflation scenarios \cite{Enqvist:2019jkb, Bae:2017tll,  Allahverdi:2020bys}.

In this work, we assume the moduli is a simple matter  approximated by $\rho_{\rm m} \propto a^{-3}$  as a baseline model of thermal inflation. The moduli field oscillates around the minimum affecting perturbations during its generation and domination, but its detailed dynamics could not be captured in our assumption. 
We leave full consideration of moduli dynamics and its observational implications as a future work.

\section*{Acknowledgements}
The authors thank Ewan Stewart for his helpful discussion and advice. 
HZ thanks Emre Onur Kahya, and the Department of Physics Engineering at Istanbul Technical University and Korea Astronomy and Space Science Institute for the hospitality. This work was supported by the DGIST-UGRP grant. SEH was 
supported by the project ``Understanding Dark Universe Using Large Scale Structure of the Universe'', funded by the Ministry of Science.

\bibliographystyle{apsrev4-2}
\bibliography{TImodel}

\end{document}